\documentclass[pra,twocolumn,superscriptaddress,showpacs]{revtex4}
\usepackage[english]{babel}
\usepackage[dvips]{graphicx}
\usepackage{amssymb}
\usepackage{amsmath}
\usepackage{color}
\begin{document}

\title{ Strong coupling of spin qubits to a transmission line resonator }

\author{Pei-Qing Jin}
\affiliation{Institut f\"ur Theoretische Festk\"orperphysik,
      Karlsruhe Institute of Technology, 76128 Karlsruhe, Germany}
\author{Michael Marthaler}
\affiliation{Institut f\"ur Theoretische Festk\"orperphysik,
      Karlsruhe Institute of Technology, 76128 Karlsruhe, Germany}
\author{Alexander Shnirman}
\affiliation{Institut f\"{u}r Theorie der Kondensierten Materie,
      Karlsruhe Institute of Technology, 76128 Karlsruhe, Germany}
\affiliation{DFG Center for Functional Nanostructures (CFN),
      Karlsruhe Institute of Technology, 76128 Karlsruhe, Germany}
\author{Gerd Sch\"on}
\affiliation{Institut f\"ur Theoretische Festk\"orperphysik,
      Karlsruhe Institute of Technology, 76128 Karlsruhe, Germany}
\affiliation{DFG Center for Functional Nanostructures (CFN),
      Karlsruhe Institute of Technology, 76128 Karlsruhe, Germany}

\date{\today}

\begin{abstract}
We propose a mechanism for coupling spin qubits formed
in double quantum dots to a superconducting transmission line resonator.
Coupling the resonator to the gate controlling the interdot tunneling
creates a spin qubit--resonator interaction with strength of tens of MHz.
This mechanism allows operating the system at a symmetry point 
where decoherence due to charge noise is minimized.
The transmission line can serve as shuttle, allowing for fast two-qubit operations
including the generation of qubit-qubit entanglement
and the implementation of a controlled-phase gate.
\end{abstract}

\pacs{03.67.Lx,73.21.La,42.50.Pq}

\maketitle

\emph{Introduction.}\textemdash
Mesoscopic electronic circuits can realize artificial quantum two-level systems
with tunable parameters, which makes them promising devices
for quantum information processing.
Among them are spin qubits formed by electron spins in quantum dots~\cite{Loss98}.
Coherent manipulations of such spin qubits have been demonstrated \cite{Petta05, Koppens, Nowack},
however, generating a non-local qubit-qubit interaction remains a challenge.
Circuit quantum electrodynamics (QED) setups \cite{Wallraff04},
with superconducting qubits coupled via a  transmission line, have been demonstrated to
provide solutions for this task~\cite{Schoelkopf,Majer}.
Stimulated by this success proposals for coupling spin qubits
to a superconducting resonator
were put forward \cite{Lukin04,Burkard06,Trif,Cottet,DR},
and experimental progress has been made
towards coupling quantum dots to a superconducting resonator \cite{DR1,DR2,DR3}.
Magnetic coupling between a resonator and a spin ensemble
was reported recently \cite{Schuster10,Bushev},
but coupling to a single spin with tiny magnetic moment remains difficult.

A strategy to increase the coupling strength is to involve charge degrees of freedom.
For spin qubits defined by singlet and triplet states in double quantum dots,
a strong coupling mechanism has been  proposed based on
transitions between singly and doubly occupied states~\cite{Burkard06}.
It requires the system to be operated away from the charge degeneracy point
with a detuning of the dot levels.
Unfortunately, the strong coupling achieved in this way
is inevitably accompanied by fast dephasing,
limiting the coherence time to the regime of nanoseconds~\cite{Petersson,SM}.

Here we propose a coupling mechanism that allows 
the system to be operated at the \emph{charge degeneracy point},
thus minimizing the effect of charge fluctuations.
The resonator couples to the gate controlling the interdot tunneling.
Its electric field, induced by vacuum fluctuations or controlled excitation,
modifies the exchange splitting between the singlet and triplet states.
In combination with an inhomogeneous Overhauser field due to
nuclear spins in the quantum dots,
both transverse and longitudinal spin-resonator coupling
(in the qubit's eigenbasis) can be achieved,
with strength controlled via a magnetic field or local electric gates.
This enables various mechanisms for two-qubit operations, with efficiencies
depending on the parameter regime.
With additional driving on the oscillator, a blue-sideband transition is
available as a strong first-order process.
In this way, fast entanglement between distant spin qubits can be achieved.

\begin{figure}[t]
\centering
\includegraphics[width=1.0\columnwidth]{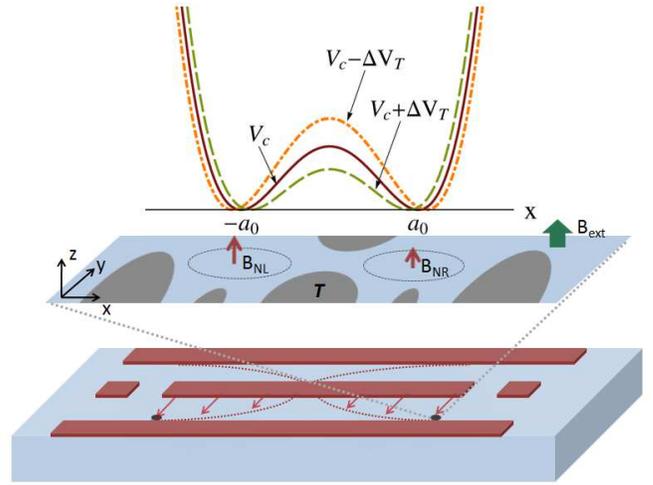}
\caption{(Color online)
Quantum dot-resonator circuit.
A spin qubit formed in a double quantum dot, each dot containing one electron,
is placed at a maximum of the electric field
inside a superconducting transmission line resonator.
The resonator electric field couples to the interdot tunnel gate T,
which modifies the tunnel barrier height.
The electrons in the dots experience a magnetic field
given by an applied field $\bf B_{\rm ext}$
and Overhauser fields $\bf B_{\rm NL/NR}$ due to nuclear spins, which are different for the two dots.
The transmission line provides the coupling to a second spin qubit 
indicated on the left.
}
\label{fig:DotRes}
\end{figure}
\emph{Model.}\textemdash
We consider a gated double quantum dot in a 2-dimensional electron gas tuned to degeneracy
as shown in Fig.~\ref{fig:DotRes}.
Following Ref.~\onlinecite{Burkard99} we assume for  definiteness that the
confining potential is
\begin{eqnarray}
 V_{\rm c} (x,y) = \frac{m\omega_0^2}{2}
 \left[ \frac{1}{4 a_0^2} ( x^2 - a_0^2 )^2 + y^2 \right].
\end{eqnarray}
The two dots, located at $\mathbf r_{\pm} = (\pm a_0, 0)$,
are separated by a parabolic tunnel barrier,
\begin{eqnarray}\label{eq:tunnelbarrier}
 V_{\rm c} (x,0) \approx V_{\rm h}  -\frac{m\omega_0^2}{4} x^2\,,
\end{eqnarray}
with height $V_{\rm h} = m\omega_0^2 a_0^2/8$ which
can be controlled by the voltage applied on the tunnel gate T.

For strong on-site Coulomb energy,
the relevant charge configuration at low temperature has one electron in each dot.
An external magnetic field $\mathbf B_{\rm ext} = B \hat z$
splits off the two spin triplet states with $m_s=\pm 1$,
which allows us to focus on the subspace spanned by the remaining triplet state
$|T_0\rangle = |-\rangle \otimes (|\!\!\uparrow\downarrow\rangle + |\!\!\downarrow\uparrow\rangle)/\sqrt{2}$
and the singlet
$|S\rangle = |+\rangle \otimes (|\!\!\uparrow\downarrow\rangle - |\!\!\downarrow\uparrow\rangle)/\sqrt{2}$.
Here $|\mp\rangle$ denote the orbitals of the triplet and singlet states.
In this two-state subspace, the double dot system is described by
\begin{eqnarray}\label{eq:Hamiltonian_st}
 H_{\rm d} = \frac{J_0}{2} \tau_z + \frac{\Delta h}{2} \tau_x \, .
\end{eqnarray}
The exchange splitting is $J_0$, and $\Delta h$ accounts for a Zeeman splitting
difference between the two dots, e.g., due to inhomogeneous nuclear spin fields \cite{Hanson}
or generated by a micromagnet attached to double quantum dots \cite{MicroM}.
The Pauli matrices are defined as
$\tau_z = |T_0\rangle\langle T_0|-|S\rangle\langle S|$
and
$\tau_x = |T_0\rangle\langle S| + |S\rangle\langle T_0|$.
Spin-orbit coupling is assumed to be weak and is not included here.

\begin{figure}[t]
 \centering
 \includegraphics[width=0.4\textwidth]{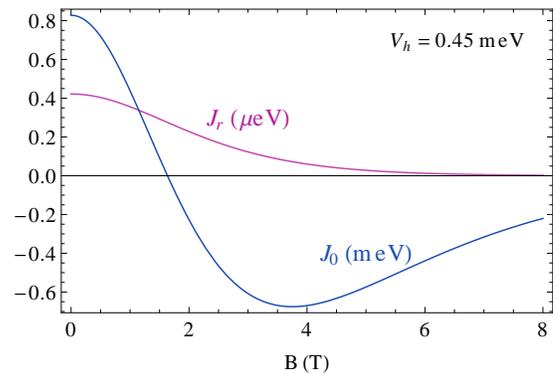}
 \caption{(Color online) Bare exchange splitting $J_0$
 and its resonator-induced part $J_{\rm r}$
 as functions of magnetic field.
 The bare exchange splitting changes sign at magnetic field $B \sim 1{\rm T}$
 because of the competition between kinetic and Coulomb energies.
 The parameters are chosen appropriate for GaAs quantum dots, as used
 in experiments \cite{Petta05,Koppens,Nowack},
 with confining potential $\hbar \omega_0 = 4.5 {\rm meV}$,
 Zeeman splitting difference $\Delta h = 1~\mu {\rm eV}$,
 and resonator-induced voltage drop $V_{\rm r} = 1\mu {\rm V}$.
 } \label{fig:JMag}
\end{figure}

An  estimate of the bare exchange splitting $J_0$
is provided by the Heitler-London model \cite{Burkard99}.
In this approach, the orbitals of the symmetric and antisymmetric two-electron states
are constructed by single-electron ground states $|L/R\rangle$
localized in the left/right quantum dots, namely,
\begin{eqnarray}
 |\pm\rangle &=& \frac{|L_1R_2\rangle \pm |L_2R_1\rangle}{\sqrt{2(1 \pm s^2)}}.
\end{eqnarray}
Here $s=\langle L|R\rangle$ denotes the overlap between the ground-states wavefunctions,
and the subscripts are introduced to label the electrons.
As shown in Fig.~\ref{fig:JMag},
in the presence of strong magnetic field,
the bare exchange splitting undergoes a sign change
due to the competition between kinetic energy and Coulomb repulsion~\cite{Burkard99,Hu,Calderon}.

We assume that a superconducting transmission line,
modeled as harmonic oscillator with frequency $\omega_{\rm r}$,
is coupled to the interdot tunnel gate T
(indicated in Fig.~\ref{fig:DotRes}).
The resonator has a significant vacuum-fluctuations-induced voltage $V_{\rm r}$
between its central conductor and the ground plane,
typically of order of $\mu V$ \cite{Wallraff04,Blais04}.
Adding this voltage to the interdot tunnel gate changes the tunnel barrier
as illustrated in Fig.~\ref{fig:DotRes}.
As long as the potential remains symmetric (see below for a discussion of possible deviations)
we can model the resonator- induced change of the tunnel barrier by
\begin{eqnarray}\label{eq:changetunnelbarrier}
 \Delta V_{\rm T} = eV_ {\rm r}\, x^2/a_0^2.
\end{eqnarray}
The exchange splitting is modified accordingly, leading to the
qubit-resonator interaction
\begin{eqnarray}
 H_{\rm c} = J_{\rm r} \tau_z (a^\dag + a^{}),
\end{eqnarray}
with the resonator-induced exchange splitting given by
\begin{eqnarray}\label{eq:exchange splitting}
 J_{\rm r}  &=& \frac{1}{2} \sum_{i=1,2}
   \Bigl [\langle T_0| \Delta V_{\rm T} (x_i) |T_0\rangle
 - \langle S| \Delta V_{\rm T} (x_i) |S\rangle \Bigr ]
                   \nonumber \\[2mm]
  &=&  e V_{\rm r}/\text{sinh} \left[ \frac{ 16 V_{\rm h} (\omega_0^2+2\, \omega_{\rm L}^2) }
  {\hbar \,\omega_0^2 \sqrt{\omega_0^2+\omega_{\rm L}^2} } \right] .
\end{eqnarray}
Here $a^\dag$ denotes the creation operator for
the resonator radiation field,
$\omega_{\rm L} = eB/2m$ is the Larmor frequency,
and $a_{\rm B} = \sqrt{\hbar/m\omega_0}$ an effective Bohr radius determined by the confinement.
The resonator-induced exchange splitting $J_{\rm r}$
increases with the wavefunction overlap.
The magnetic field compresses the electron orbitals, and hence
$J_{\rm r}$ decreases, as shown in Fig.~\ref{fig:JMag}.

The coupling mechanism proposed here
can be realized in experiments by fabricating a finger-shaped electric gate
extending from the resonator to the interdot tunnel gate.
A similar setup was used in the experiments of Ref.~\onlinecite{DR2}
to couple charge states of a double quantum dot to the resonator.
In realistic situations,
the finger-shaped gate may be asymmetric with respect to
the left and right dots, adding a small asymmetric contribution to $\Delta V_{\rm T}$.
In addition, the resonator voltage $V_{\rm r}$ could also couple to other 
gates controlling the confinement.
However, these effects are weak compared to the confining energy $\hbar \omega_0$ of order of meV
and do not change the charge configuration (one electron per dot).
Furthermore, modifications that are odd in $x$ have vanishing matrix elements, 
$\langle S/T_0| x_1^n +x_2^n |S/T_0 \rangle = 0$ for n being odd.
I.e., our coupling scheme is insensitive to small odd variations, while
an additional even variation  modifies our results only quantitatively.

\emph{Coupling strength.}\textemdash
In the eigenbasis $\{|E\rangle, |G\rangle \}$ of the qubit Hamiltonian~(\ref{eq:Hamiltonian_st}),
both transverse and longitudinal coupling
between the double dot and resonator arise,
\begin{eqnarray}\label{eq:HDR}
 H_{\rm qr} &\!\!=\!\!& \frac{\hbar \omega_{\rm q}}{2} \sigma_z
       + \hbar \omega_{\rm r} a^\dag a^{}
       + \hbar(g_x \sigma_x + g_z \sigma_z)(a^\dag+a^{}),
\end{eqnarray}
with coupling strengths $g_x = -J_{\rm r} \sin\theta/\hbar $ and $g_z=J_{\rm r} \cos\theta/\hbar $
depending on the mixing angle $\theta = \arctan (\Delta h/J_0)$.
Here, the Pauli matrices $\sigma_i$ are defined in the qubit eigenbasis,
e.g., $\sigma_z =|E\rangle\langle E|-|G\rangle\langle G|$,
and the qubit splitting is $\omega_{\rm q} = \sqrt{J_0^2 + \Delta h^2}/\hbar$.

The transverse coupling allows energy exchange between qubit and resonator.
It is maximized when the bare exchange splitting vanishes ($\theta = \pi/2$),
in which case the eigenstates of the qubit are simply those with spin configurations
$|\!\!\uparrow\downarrow\rangle$ and $|\!\!\downarrow\uparrow\rangle$.
The strength of the transverse coupling can reach several tens of MHz,
given that the electrostatic potential induced by the resonator
is of order of $\mu$eV.
If the spin qubit reaches the expected long decoherence time, $T_2 \sim 10 \, \mu s$,
and the superconducting transmission line resonator
a  decay rate $\kappa/2\pi\sim 100 \, {\rm kHz}$,
the system reaches the strong coupling regime.
The actual coupling can be even stronger
since the Heitler-London approach underestimates the exchange splitting.
To stay within the validity regime of the Heitler-London approach~\cite{Calderon},
we presented here results for a relatively strong confining potential
$\hbar\omega_0=4.5 \, {\rm meV}$ and lower bound of interdot distance $ a_0 \gtrsim 0.9 \, a_{\rm B}$.
For weaker confining potential the coupling strength will be higher.

An important property of the proposed dot-resonator system
is the existence of a longitudinal coupling.
Many superconducting qubits have almost no longitudinal
coupling, or the coupling vanishes at the degeneracy point
where dephasing effects are minimized.
In the dot-resonator system, charge fluctuation induced dephasing is minimized
by involving only states with the same charge configuration (one electron in each dot).
This, however, does not switch off the longitudinal coupling.
Actually, a strong longitudinal coupling of hundreds of MHz is possible
when the bare exchange splitting $J_0$ dominates over the Zeeman splitting difference $\Delta h$.

\emph{Two-qubit gates.}\textemdash
We consider two distant qubits ($i=1,2$)
coupled to a common resonator mode,
\begin{eqnarray}
 H_{\rm 2q} \!\! &=& \!\!
 \sum_{i=1,2} \frac{\hbar\, \omega^{(i)}_{\rm q}}{2} \sigma^{(i)}_z
 +\hbar\, \omega_{\rm r}\, a^\dag a^{}
                    \nonumber \\
 &+&  \sum_{i=1,2} \hbar \left[ g^{(i)}_{x}\sigma^{(i)}_x+g^{(i)}_{z}\sigma^{(i)}_z \right] (a^\dag+a^{}).
\end{eqnarray}
Given the strong coupling between spin qubits and the transmission line
it is possible to perform controlled two-qubit gates. As examples
we will discuss in the following: (i) the generation of qubit-qubit entanglement
via blue-sideband transitions,
(ii) the implementation of  a controlled-phase (CPhase) gate
based on a direct longitudinal qubit-qubit interaction,
and (iii) a $\sqrt{\rm iSWAP}$ gate via exchange of virtual photons
in the resonator.

(i)
The blue-sideband transition, which excites the qubit $i$ and the resonator simultaneously,
is induced by resonant interaction of the form
\begin{eqnarray}\label{eq:BST}
 H^{(i)}_{\rm BST} = \hbar\, \Omega^{(i)}_{\rm BST}
 \left[ a^\dag \sigma^{(i)}_+ + a^{}\sigma^{(i)}_- \right].
\end{eqnarray}
If the system is initialized in the ground state $|G^{(i)}\rangle\otimes|0\rangle$,
turning on the interaction $H^{(i)}_{\rm BST}$ for a period $\tau = \pi/2\,\Omega^{(i)}_{\rm BST}$
creates an entangled qubit-resonator state
$(|G^{(i)}\rangle\otimes|0\rangle+|E^{(i)}\rangle\otimes|1\rangle)/\sqrt{2}$.
Applying this procedure to entangle each qubit with the resonator separately
allows the generation of qubit-qubit entanglement \cite{BST,Wallraff0709}.
This scheme has the advantage that the generation of entanglement is fast when the resonator is strongly driven,
and no tuning of the qubit frequencies is required.

For the dot-resonator system, the blue-sideband transition
is achieved by driving the resonator with amplitude $\epsilon_d$ and
with frequency
$\omega_{\rm d} = \omega_{\rm q}+\omega_{\rm r}$.
(Here the qubit label is omitted for simplicity.)
With a driving field amplitude $\epsilon_{\rm d} \ll \omega_{\rm q}\,\omega_{\rm r}/g_x$,
which is realistic for a typical circuit QED setup,
the Rabi frequency reduces to~\cite{Blais07}
\begin{eqnarray}
\Omega_{\rm BST} = \frac{2\, \epsilon_{\rm d} \, g_x \, g_z }
 { \omega_{\rm q}\, \omega_{\rm r} } .
\end{eqnarray}
With a driving field amplitude of the order of hundred MHz,
the magnitude of the Rabi frequency $\Omega_{\rm BST}$ can reach several tens of MHz.
This strong Rabi frequency relies on the existence of the longitudinal coupling
in the dot-resonator system.
Without  longitudinal coupling,
the blue side-band transition is only accessible in a second-order process,
since such a system is invariant under a parity transformation
with the operator $P=\exp{(-i\pi a^\dag a^{})}\sigma_z$,
while the driving responsible for blue-sideband transitions is of odd parity \cite{Blais07}.
The longitudinal coupling breaks the symmetry of the system,
allowing the blue-sideband transition in first order.

\begin{figure}[t]
\centering
\includegraphics[width=0.7\columnwidth]{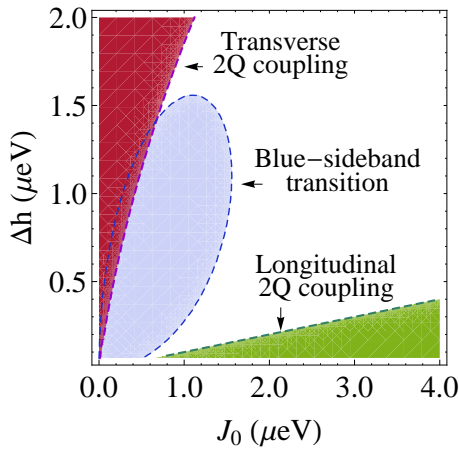}
\caption{(Color online) Parameter space
for two-qubit (2Q) interactions.
Longitudinal and transverse 2Q couplings,
as well as blue-sideband transitions become strong
(exceeding 10 MHz) in the green, red, and blue areas, respectively.
The resonator-induced part of the exchange splitting
is assumed to be $J_{\rm r} = 0.3 \, \mu {\rm eV}$,
the resonator frequency is $\omega_{\rm r}/2\pi = 3\, {\rm GHz}$,
and the driving field strength is $\epsilon_{\rm d}/2\pi = 450\, {\rm MHz}$.
}
\label{fig:PD}
\end{figure}
(ii) When the Zeeman splitting difference $\Delta h^{(i)}$ is negligible
compared to the bare exchange splitting $J_0^{(i)}$,
a direct longitudinal qubit-qubit interaction arises.
After the unitary transformation with
$ U = \exp \left[(a^\dag-a^{}) \sum_{i=1,2} g^{(i)}_z \sigma^{(i)}_z/\omega_{\rm r}\right]$,
the effective Hamiltonian $H_{\rm zz} = U H_{\rm 2q} U^\dag$ is given by
\begin{eqnarray}
 H_{\rm zz} &=& \sum_{i=1,2} \frac{\hbar\, \omega^{(i)}_{\rm q}}{2} \sigma^{(i)}_z
 -  \frac{2 J^{(1)}_{\rm r} J^{(2)}_{\rm r}}{\hbar\, \omega_{\rm r}} \sigma^{(1)}_z \sigma^{(2)}_z
                    \nonumber \\
& & + \, \hbar\,\omega_{\rm r}\, a^\dag a
.
\end{eqnarray}
In the considered parameter regime, $J_0^{(i)} \gg \Delta h^{(i)}$,
the resonator-induced exchange splitting $J_{\rm r}^{(i)}$ can reach hundreds of MHz,
which leads to a strong longitudinal qubit-qubit coupling of tens of MHz.
It allows realizing efficiently a CPhase gate.

(iii) A strong transverse qubit-resonator coupling allows for a fast two-qubit operation
in the dispersive regime \cite{Blais04}.
In this regime, two qubits are far-detuned from the resonator but in resonance with each other.
Their interaction is mediated by the exchange of virtual photons.
In second order perturbation theory, the effective transverse interaction becomes \cite{TC,Imamoglu99}
\begin{eqnarray}
 H_{\rm DIS} = \hbar\, \Omega_{\rm DIS}
 (\sigma_+^{(1)} \sigma_-^{(2)} +\sigma_-^{(1)} \sigma_+^{(2)} ),
\end{eqnarray}
with coupling strength 
\begin{eqnarray}
 \Omega_{\rm DIS} = g_x^2/( \omega_{\rm q} - \omega_{\rm r})
\end{eqnarray}
which can reach several MHz for a qubit-resonator detuning of hundreds of MHz.
(For simplicity we assumed the two qubits to be identical.)

\emph{Discussion and Summary.}\textemdash
In circuit QED setups with superconducting qubits,
switching off the qubit-qubit interaction is usually achieved
by tuning the two qubits out of resonance.
For the dot-resonator system,
tuning qubit frequencies is always accompanied with changing the qubit-resonator coupling,
since both depend on the interdot tunnel barrier.
By increasing, e.g., the tunnel barrier height of one double dot
one increases the frequency detuning between the qubits
and at the same time reduces the qubit-resonator coupling.
As a result the qubit-qubit interaction in the dot-resonator system
can be switched off highly efficiently.

We summarize the scenario of the two-qubit interactions in Fig.~\ref{fig:PD}.
The parameter space is spanned by
the bare exchange splitting $J_0$ and the Zeeman splitting difference $\Delta h$.
They are the key elements for the spin qubit and can be measured in experiments \cite{Petta05}.
For simplicity we assume the resonator-induced part of the exchange splitting to be constant,
$ J_{\rm r} =0.3 \, {\rm \mu eV}$,
since in the considered parameter region it varies by less than $ 0.01\, {\rm \mu eV}$.
The colored areas indicate the regions
where, for realistic parameters, the corresponding qubit-qubit interaction is stronger than 10 MHz.
For strong bare exchange splitting,
a longitudinal qubit-qubit coupling of several tens of MHz can be reached, allowing
for an efficient CPhase gate.
When the Zeeman splitting dominates 
the transverse qubit-qubit interaction is strong,
which allows for the implementation of an $\rm{\sqrt{iSWAP}}$ gate.
With comparable bare exchange splitting and Zeeman splitting difference,
strong blue-sideband transitions are favorable to produce fast qubit-qubit entanglement.

In the examples presented above we used parameters appropriate for the 
experiments performed with GaAs samples.
Another promising material for quantum dots is Si,
for which, due to the weak hyperfine interaction, spin qubits have been shown to have a 
much longer dephasing time
$T_2^* \sim 360~\rm{ns}$ \cite{SiQD}.
The coupling mechanism proposed here also applies to Si quantum dots
(provided the valley degeneracy is lifted by a splitting of several meV \cite{SVA}),
except for quantitative changes due to the larger effective mass and smaller dielectric constant.

We acknowledge helpful discussions
with I. Kamleitner, J.H. Cole,  A. Romito, and J. Weis,
and the support from the Baden-W\"{u}rttemberg Stiftung
via the 'Kompetenznetz Funktionelle Nanostrukturen' and the DFG via the 
Priority Program 'Semiconductor Spintronics'.

\end{document}